\documentclass[newstyle,twocolumn,proceedings]{rmaa}
\newcommand{\ia}{\'{\i}}

\newcommand{\zsun}{$Z_{\odot}$}

\title{Modeling Stellar Populations in Star Clusters and Galaxies
}
\author{Gustavo Bruzual A. \affil{CIDA, M\'erida, Venezuela} }
\fulladdresses{
\item Gustavo Bruzual A.: Centro de Investigaciones de Astronom\'\i a,
Apartado Postal 264, M\'erida 5101-A, Venezuela (bruzual@cida.ve).}

\shortauthor{G. Bruzual}
\shorttitle{Modeling Stellar Populations}

\abstract{
Combining population synthesis models with simple Montecarlo simulations of
stochastic effects in the number of stars occupying sparsely populated stellar
evolutionary phases in the HRD, I show that the scatter observed in the
photometric magnitudes and colors of LMC and NGC 7252 star clusters can be
understood in the framework of current stellar evolution theory.
The use of a high resolution stellar spectral atlas in population synthesis
models improves considerably the quality of the fits to observed galaxy SEDs,
making the assignment of a spectroscopic age to stellar populations more
reliable than with low spectral resolution models.
}


\resumen{
Combinando modelos de s\ia ntesis de poblaciones estelares con simulaciones
Montecarlo de los efectos estoc\'asticos en el n\'umero de estrellas que
ocupan estadios evolutivos escasamente poblados en el DHR, se concluye que
la dispersi\'on observada en la magnitud y colores fotom\'etricos de
c\'umulos estelares en la NGM y NGC 7252 pueden entenderse en el marco de
la teor\ia a actual de evoluci\'on estelar. El uso de bibliotecas de 
espectros estelares de alta resoluci\'on en los modelos de s\ia ntesis de
poblaciones mejora considerablemente la calidad de los ajustes a las
distribuciones espectrales de energ\ia a observadas, y permite asignar
edades espectrosc\'opicas a poblaciones estelares con mayor confiabilidad
que con los modelos de baja resoluci\'on espectral.
}

\begin{document}
\maketitle

\section{Introduction}
Evolutionary population synthesis has become an important tool to study the
stellar population content and its evolution over cosmological time scales 
in star cluster and galaxies. Bruzual (2001, 2002a) has presented a detailed
description of the population synthesis technique, including main results,
comparison with observations, and sources of uncertainties.
For reasons of space, in this paper I include a very brief summary of
population synthesis and an update on recent developments since the previous
papers. References to complementary work are provided.

\section{Population Synthesis}
The current theory of stellar evolution makes detailed predictions concerning
the time dependence of the bolometric luminosity and effective temperature of
a star of a given mass and chemical composition.
Complete sets of evolutionary tracks for stars of a wide mass range and various
initial metal contents covering all significant stellar phases are available
in the literature (e.g. Fagotto et al. 1994a, b, c; Girardi et al. 2000). 
Assuming an initial mass function (IMF, e.g. Salpeter 1955), we can compute
the number of stars of a given mass born at time zero, and then follow the
evolution of this population in the HRD using a specific set of evolutionary
tracks.
Knowledge of the spectral energy distribution (SED) at each position in the
HRD visited by the stars during their evolution, allows us to compute the
integrated SED for this initial-burst or simple stellar population (SSP) as a
function of time.
By means of a convolution integral (Bruzual \& Charlot 1993) the evolution
of the SED can be computed for an arbitrary star formation rate (SFR) and 
a chemical enrichment law.

\section{Stellar Libraries}
Several stellar spectral libraries for solar metallicity (\zsun) stars are
currently available. The Pickles (1998) atlas provides good coverage of the HRD
for stars of this metallicity at medium spectral resolution.
Le Borgne et al. (2002) have compiled an equally comprehensive atlas at
3\AA\ spectral resolution (1\AA\ sampling) which is complete for stars of
solar metallicity, but includes a large number of spectra
of non-solar metallicity stars. On the theoretical side, the compilations by
Lejeune et al. (1997, 1998) and Westera et al. (2002) provide libraries of model
atmospheres for stars of various metallicities but at $\approx$20\AA\ spectral
resolution. The last two libraries are largely based on the Kurucz (1995)
series of model stellar atmospheres. Figs 1 and 2 show the time evolution of
the SED for a \zsun\ SSP model computed for these three stellar libraries in
increasing order of spectral resolution.
Throughout this paper I use the $x=1.35$ Salpeter (1955) IMF, and for the
lower and upper mass limits of star formation, I assume $m_l=0.09$ and
$m_u=125$ M$_\odot$, respectively.
The set of population synthesis models that use the Le Borgne et al. (2002)
high resolution stellar spectral library will be discussed by Bruzual \&
Charlot (2002, BC02 hereafter).
All the models shown in Figs 1 and 2 have the same behavior in what respects to
integrated photometric properties of the stellar population, e.g. luminosity
and color evolution.
The highest resolution model reproduces in much greater detail the spectral
features of integrated stellar populations, observed, for instance,
with the high resolution spectrograph available in HST and in the new
generation of ground based high performance large optical telescopes
(Keck, VLT, Sloan).
See for example Figs 1 and 2 of Bruzual (2002c) and \S5 below.

\begin{figure}[!t]
  \includegraphics[width=\columnwidth]{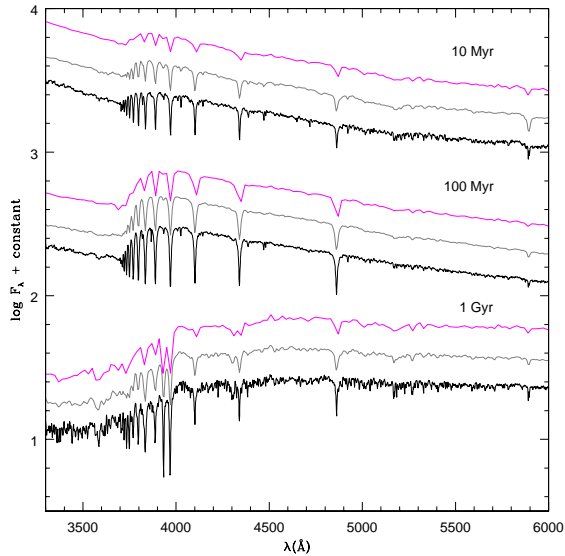}
  \caption{
  SED of a $Z =$ \zsun\ BC02 SSP model at 10, 100, and 1000 Myr computed with
  different stellar spectral libraries:
  {\it (light gray)} $\approx$20\AA\ resolution (Lejeune et al. 1997),
  {\it (medium gray)} sampling interval of 5~\AA\ pixel$^{-1}$ and resolution
  of $\lambda/\Delta\lambda\approx 500$ (Pickles 1998), and
  {\it (black)} sampling interval of 1~\AA\ pixel$^{-1}$ and resolution
  of $\lambda/\Delta\lambda\approx 1700$ (Le Borgne et al. 2002).
  The SEDs have been shifted in the vertical direction for clarity.
  }
\end{figure}

\begin{figure}[!t]
  \includegraphics[width=\columnwidth]{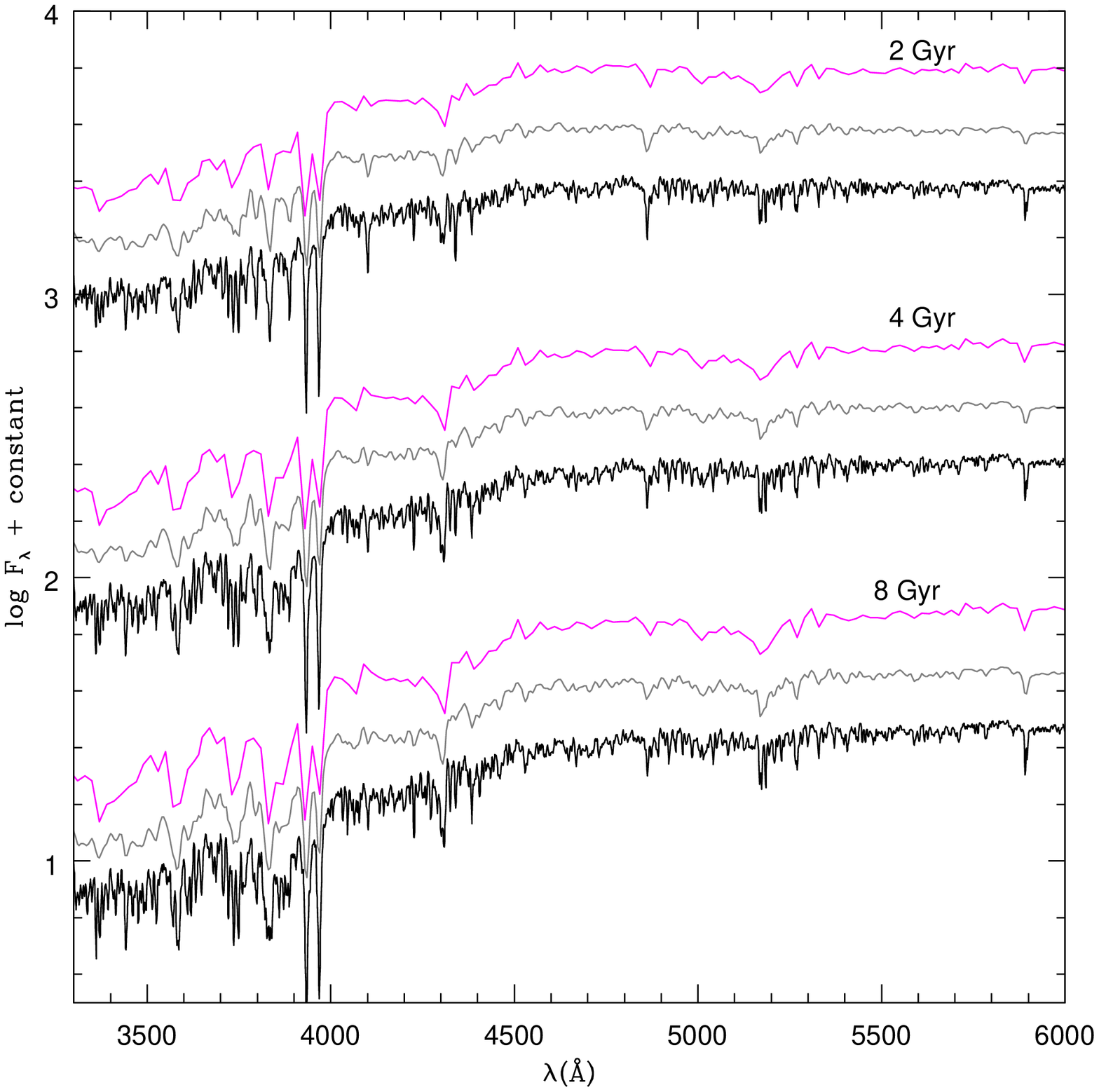}
  \caption{
  SED of a $Z =$ \zsun\ BC02 SSP model at 2, 4, and 8 Gyr computed with
  different stellar spectral libraries:
  {\it (light gray)} $\approx$20\AA\ resolution (Lejeune et al. 1997),
  {\it (medium gray)} sampling interval of 5~\AA\ pixel$^{-1}$ and resolution
  of $\lambda/\Delta\lambda\approx 500$ (Pickles 1998), and
  {\it (black)} sampling interval of 1~\AA\ pixel$^{-1}$ and resolution
  of $\lambda/\Delta\lambda\approx 1700$ (Le Borgne et al. 2002).
  }
\end{figure}

\begin{figure}[!t]
\includegraphics[width=\columnwidth]{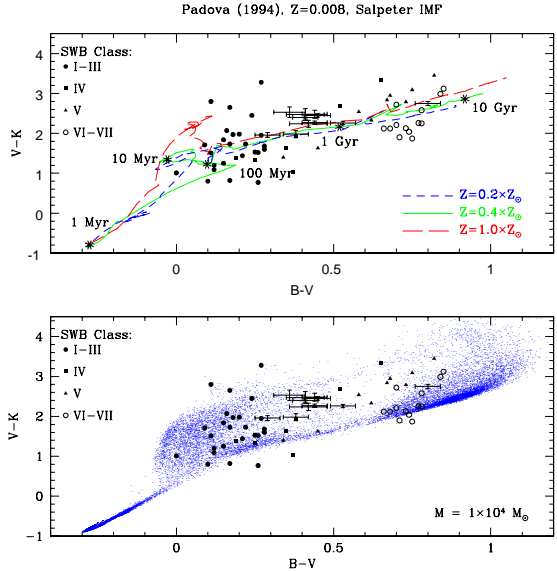}
\caption{
$V-K$ vs. $B-V$ color diagram.
The large symbols represent LMC globular clusters discriminated according to
their SWB class.
The points with error bars correspond to the star clusters in NGC 7252.
The lines represent the evolution of BC02 SSP models for various metallicities.
The dots in the bottom frame indicate the model colors when stochastic
fluctuations in the number of stars are taken into account (Bruzual 2002b).
}
\end{figure}

\begin{figure}[!t]
\includegraphics[width=\columnwidth]{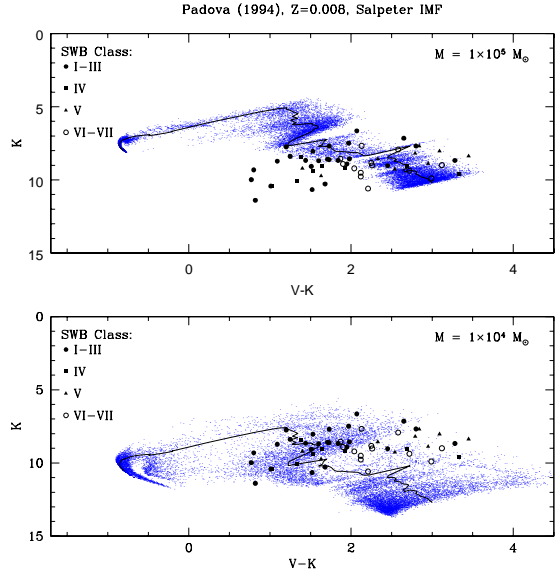}
\caption{
$K$ (at the distance of the LMC) vs. $V-K$ diagram.
The symbols represent LMC globular clusters discriminated according to
their SWB class.
The line represents the $Z = 0.4\times Z_\odot$ SSP model
with no fluctuations. The dots indicate the results of simulations including
stochastic fluctuations in the number of stars in the same model
for a  total cluster mass of $1\times10^5$ (top) and $1\times10^4$ M$_\odot$
(bottom).
}
\end{figure}

\section{Star Cluster Colors}

The differences between the observed integrated colors of star clusters and
the colors predicted for simple stellar populations by population synthesis
models are, in general, larger than allowed by observational errors.
This is particularly true for intermediate age clusters in optical-IR colors,
like $V-K$, but to a lesser extent the scatter is also large in $B-V$. 
In Fig 3 I compare the observed $B-V$ and $V-K$ colors of LMC globular cluster
(van den Bergh 1981; Persson et al. 1983, respectively) and the star clusters
in NGC 7252 (Maraston et al. 2001) with model predictions.
The lines in the top frame represent the evolution in this plane
of BC02 SSP models computed using the Padova evolutionary tracks
for $Z = 0.2\times Z_\odot,\ 0.4\times Z_\odot$, and $Z_\odot$
(Fagotto et al. 1994a,b,c), the Salpeter (1955) IMF, and the Lejeune
et al. (1997) stellar atlas.
The * symbols along the $Z = 0.4\times Z_\odot$ line mark the model colors
at the age indicated by the labels, and can be used to roughly date the
clusters.
The small dots in the bottom frame of Fig 3 indicate the $B-V$ and $V-K$ colors
resulting from different simulations by Bruzual (2002b), who has followed the
Santos \& Frogel (1997) Montecarlo technique to show that these cluster colors
are consistent with model colors, if stochastic fluctuations in the number of
stars in sparsely populated evolutionary stages are properly included into the
models.
100 simulations were run at each of 220 time steps or isochrones, obtained from
the $Z = 0.4\times Z_\odot$ SSP model shown in the top frame of the figure.
Each simulation was stopped when the total cluster mass (including dead stars)
reached $1\times10^4$ M$_\odot$.
The fluctuations in the colors become larger as the cluster mass decreases.
The model with no fluctuations is equivalent to an infinite number of stars
populating the IMF.
The expected fluctuations in $V-K$ for a $1\times10^4$ M$_\odot$ cluster
amount to almost 2 magnitudes, in close agreement with the range of
colors observed at a given age, and is considerably broader than for
the $1\times10^5$ M$_\odot$ case.
At the intermediate ages the models are redder in $V-K$ than the analytic IMF
model because of a larger number of AGB stars, which appear naturally as a
consequence of stochastic fluctuations in the IMF.
The region in this plane occupied by the observations and
the cluster simulations overlap quite well.
This comparison seems to favor a not very large mass for these clusters,
if stochastic fluctuations are responsible of the color variations.
This conclusion is supported by Fig 4 in which I plot $K$ (at the distance
of the LMC) vs. $V-K$. The region covered by the $1\times10^4$ M$_\odot$ cluster
simulations matches the region covered by the observations very well.
See Bruzual (2002b) and references therein for details and Cervi\~no et al.
(2000) for another approach to this problem.

\begin{figure}[!t]
  \includegraphics[width=\columnwidth]{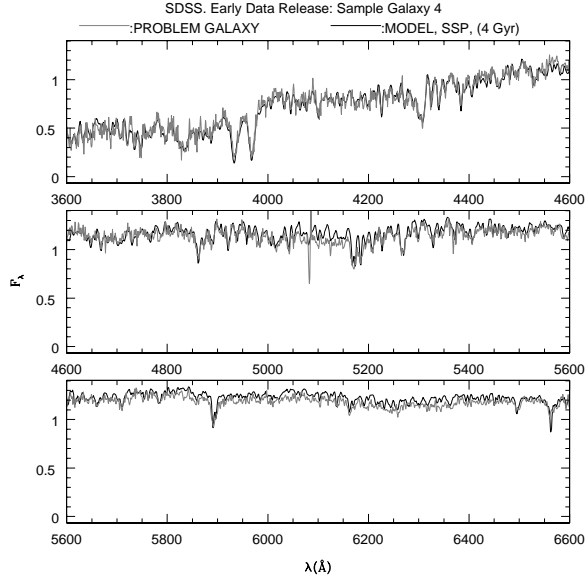}
  \caption{
  Comparison of a galaxy SED from the SDSS early data release (gray line)
  and the 4 Gyr, $Z =$ \zsun\ BC02 model SSP computed for the Salpeter (1955)
  IMF using the Le Borgne et al. (2002) stellar atlas (black line).
  }
\end{figure}

\begin{figure}[!t]
  \includegraphics[width=\columnwidth]{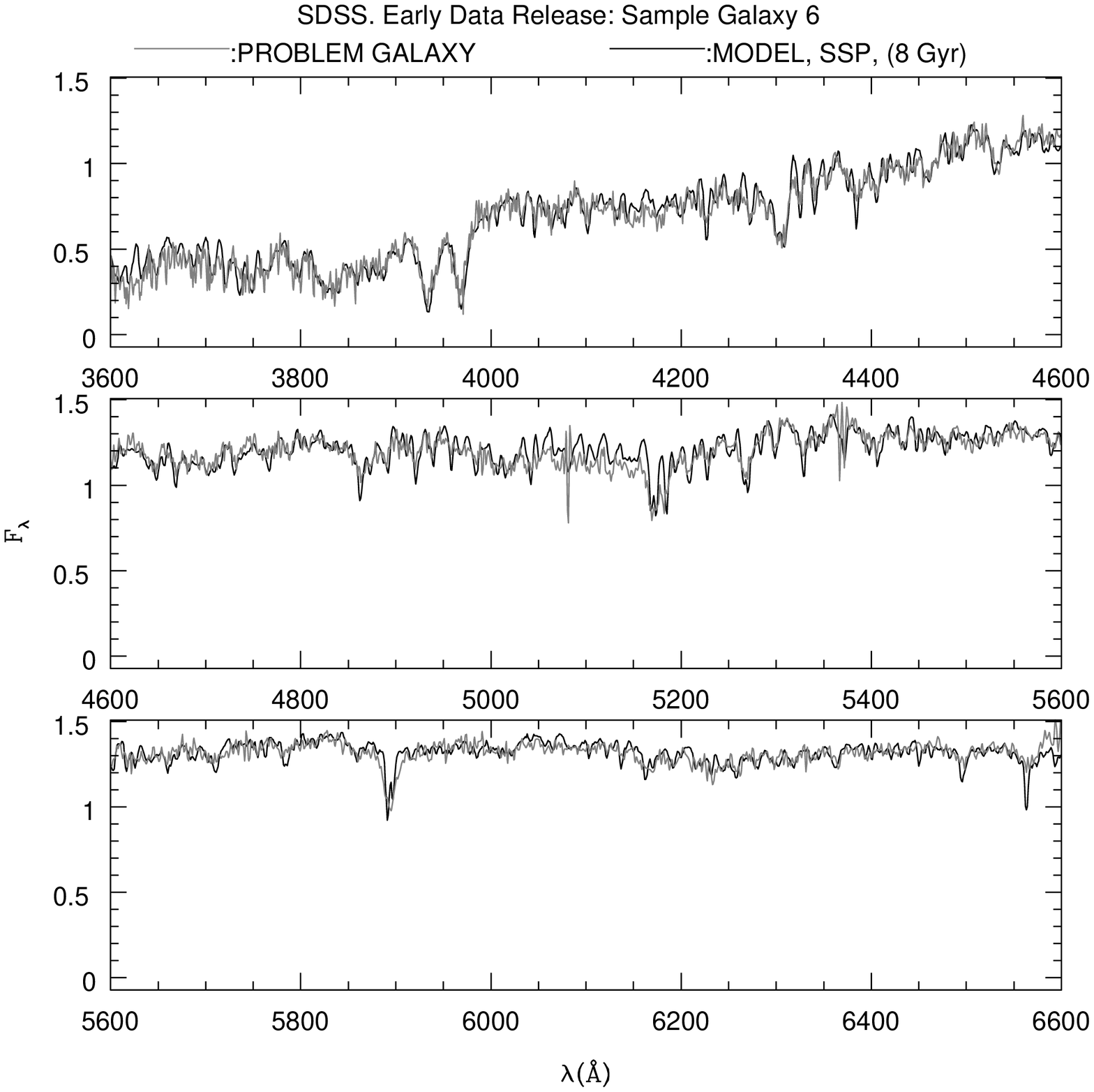}
  \caption{
  Comparison of a galaxy SED from the SDSS early data release (gray line)
  and the 8 Gyr, $Z =$ \zsun\ BC02 model SSP computed for the Salpeter (1955)
  IMF using the Le Borgne et al. (2002) stellar atlas (black line).
  }
\end{figure}

\begin{figure}
  \includegraphics[width=\columnwidth]{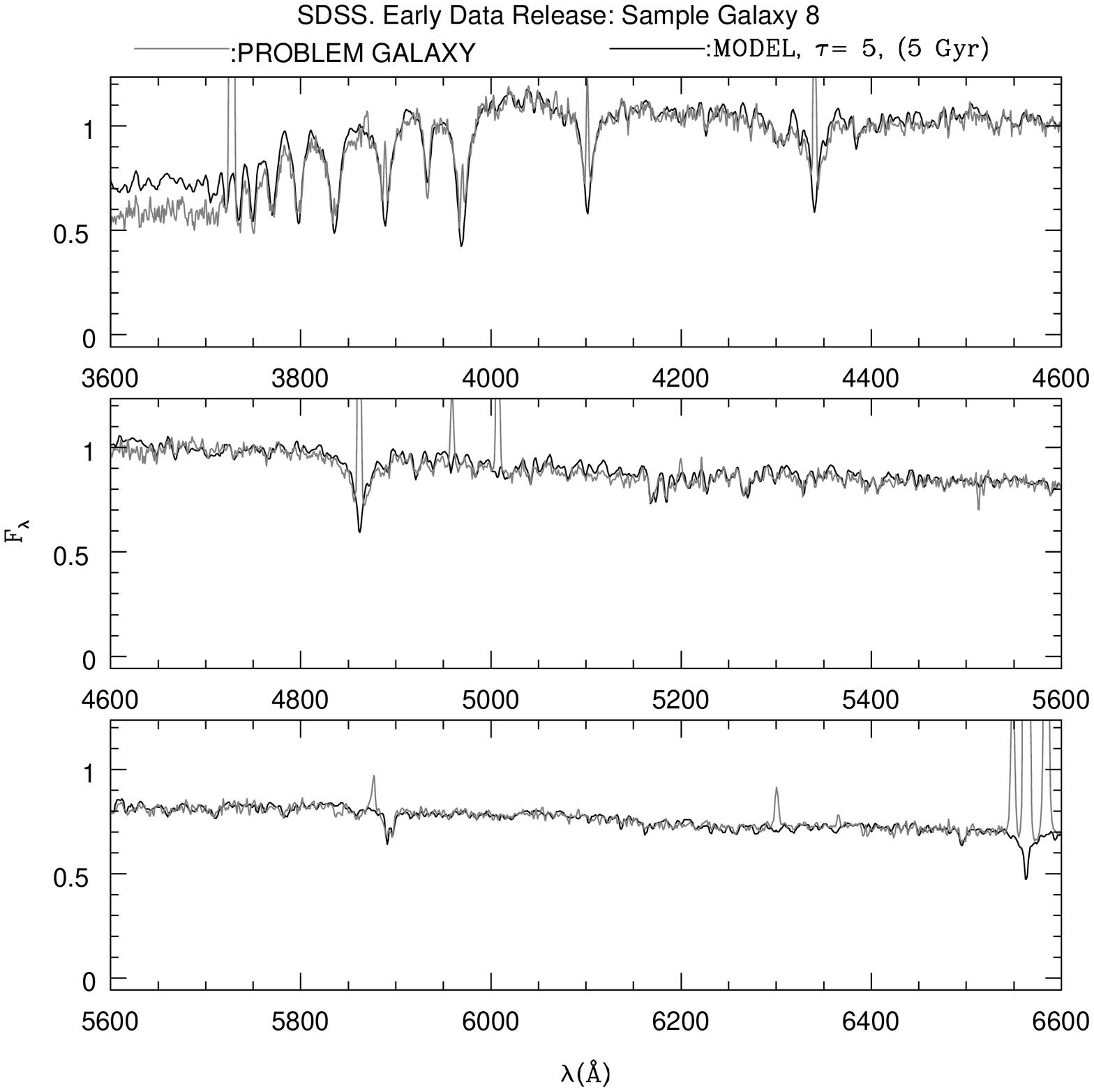}
  \caption{
  Comparison of a galaxy SED from the SDSS early data release (gray line)
  and the 5 Gyr, $Z =$ \zsun\ BC02 model
  with star formation rate $\Psi(t) \propto \exp(-t/\tau)$ with $\tau = 5$ Gyr
  (black line).
  }
\end{figure}

\begin{figure}[!t]
  \includegraphics[width=\columnwidth]{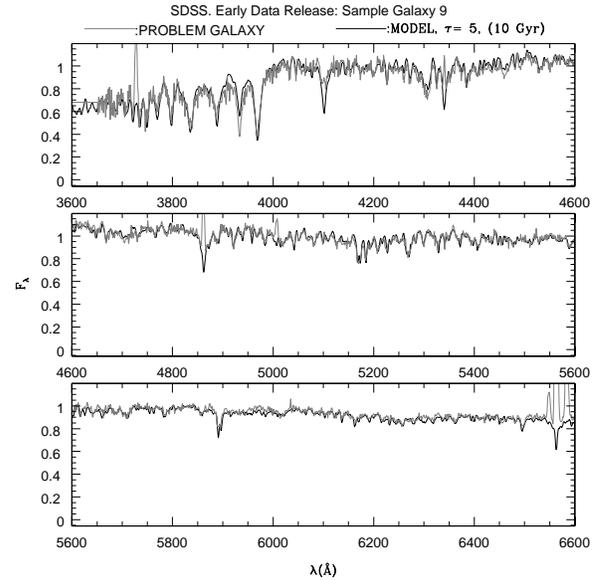}
  \caption{
  Comparison of a galaxy SED from the SDSS early data release (gray line)
  and the 10 Gyr, $Z =$ \zsun\ BC02 model
  with star formation rate $\Psi(t) \propto \exp(-t/\tau)$ with $\tau = 5$ Gyr
  (black line).
  }
\end{figure}

\section{Modeling Galaxy SEDs}

Using the BC02 models we can derive the spectroscopic age that
characterizes a given observed SED.
A model consists of a spectrum which evolves in time from age zero to 20
Gyr in 221 time steps. By minimizing
$\chi^2 = \sum [ log~ F_\lambda(observed) - log~ F_\lambda(model)]^2$,
we derive for each problem SED the (spectroscopic) age at which the model
matches more closely the observed spectrum.
Figs 5 to 8 compare the observed SEDs of 4 galaxies in the SDSS early
data release with the best fitting BC02 model SED.
Figs 5 and 6 correspond to relatively old stellar populations, which are
reproduced remarkably well by a SSP at age 4 and 8 Gyr, respectively.
On the other hand, Figs 7 and 8 correspond to galaxies which are actively
forming stars, as indicated by the emission lines seen in their SEDs.
The nebular lines emitted by ionized gas surrounding hot stars are not 
included in the BC02 models (see Charlot \& Longhetti 2001; Magris et al. 2002)
but the fit to the continuum and the absorption lines in the observed spectra
is otherwise excellent. By extending this analysis to intermediate and high
redshift galaxies,
Bruzual (2002c) has shown how the SEDs and/or multi band photometry of galaxies
from $z=0$ to $z=4$ can be ordered in a plausible sequence, along
which the spectra of galaxies may evolve when passive evolution dominates
the evolution of simple stellar populations.

\section{Discussion}

In this paper I have briefly summarized the results of simple numerical 
simulations which show clearly how the range of colors observed in intermediate
age star clusters can be understood on the basis of current stellar evolution
theory, if we take into account properly the stochastic fluctuations expected
in the number of stars occupying sparsely populated evolutionary stages in the
HRD. There does not seem to be any need to introduce ad-hoc assumptions in
population synthesis models, which translate into a departure from our current
understanding of stellar evolution theory, in order to explain the observed
range of values of cluster colors and magnitudes, as suggested by Maraston
et al. (2001).
The conclusion that cluster masses in the range around $1\times10^4$ M$_\odot$
are preferred, as well as the dependence of the amplitude of the fluctuations
in the photometric magnitude and
colors on this value of the cluster mass, should be explored in detail.

The use of a high resolution stellar spectral atlas in population synthesis
models improves considerably the quality of the fits to observed galaxy SEDs,
making the assignment of a spectroscopic age to stellar populations more
reliable than with low spectral resolution models.
The error in the age determination depends on the quality of the observational
data and on the uncertainties in the population synthesis models.
In general, the relative error in the spectroscopic age increases
with decreasing galaxy age. As shown by Bruzual (2002c), 
it is possible to establish a plausible sequence along
which the spectra of galaxies may evolve when passive evolution dominates
the evolution of simple stellar populations, at a rate that is consistent
with that expected in the most accepted cosmological model describing our
universe,
$H_0 = 70$ km s$^{-1}$ Mpc$^{-1}$, $\Omega_M = 0.28$, $\Omega_\Lambda = 0.72$,
$t_U(0)=13.75$ Gyr.

\adjustfinalcols
Work in progress suggests that the use of high resolution libraries for
various metallicities will allow
us to break the age-metallicity degeneracy and
to derive a reliable star formation history from a galaxy SED (Mateu et al.
2002).


\end{document}